\documentclass[a4paper,12pt]{article}
\usepackage{amsmath,amssymb,array,float,graphics}
\usepackage{pstricks,epsfig}
\usepackage{cite}
\numberwithin{equation}{section}

\newcommand{\ii}{\mathrm{i}}

\newcommand{\pd}{\partial}

\newcommand{\ket}[1]{\left|#1\right\rangle}

\newcommand{\tr}{\mathop{\mathrm{Tr}}\nolimits}

\newcommand{\Op}{\mathcal{O}}
\newcommand{\ft}[2]{{\textstyle\frac{#1}{#2}}}
\def\tilde{\widetilde}
\def\1bar{1\hskip -.275cm -}
\def\2bar{2\hskip -.275cm -}
\def\3bar{3\hskip -.275cm -}

\newsavebox{\uuunit}

\newcommand{\be}{\begin{equation}} \newcommand{\ee}{\end{equation}}
\newcommand{\bea}{\begin{eqnarray}} \newcommand{\eea}{\end{eqnarray}}
\newcommand{\ben}{\begin{displaymath}}
\newcommand{\een}{\end{displaymath}}
 
\newcommand{\nn}{\nonumber}

\newcommand{\nc}{\newcommand}
\nc{\la}{\lambda} \nc{\alf}{\alpha} \nc{\tht}{\theta}
\nc{\eps}{\epsilon} \nc{\ga}{\gamma} \nc{\Ga}{\Gamma}
\nc{\De}{\Delta} \nc{\de}{\delta} \nc{\si}{\sigma}
\nc{\ka}{\kappa} \nc{\om}{\omega} \nc{\qq}{\quad\quad}
\nc{\nf}{\infty} \nc{\dl}{\mathop{\smash{\cal L}}}
\nc{\ol}{\overline} \nc{\beq}{\begin{equation}}
\nc{\barr}{\begin{array}} \nc{\earr}{\end{array}}
\nc{\eeq}{\end{equation}} \nc{\beqa}{\begin{eqnarray}}
\nc{\dst}{\displaystyle}\nc{\pt}{\partial}
\nc{\eeqa}{\end{eqnarray}} \nc{\nnb}{\nonumber}
\nc{\bs}{\backslash}        \nc{\mbb}{\mathbb}
\nc{\brm}{\begin{remunerate}} \nc{\erm}{\end{remunerate}}
\nc{\vareps}{\varepsilon} \nc{\tb}{\tilde\beta_0} \nc{\ts}{\tilde
s} \nc{\tth}{\tilde \theta}
\newcounter{muni}
\usecounter{muni}
  \nc{\lapdec}{\mathop{\Delta}}

\newenvironment{remunerate}{\begin{list}{{\rm \arabic{muni}.}}
{\usecounter{muni}
\setlength{\leftmargin}{0pt}\setlength{\itemindent}{38pt}}}{\end{list}}
\newdimen\squaresize \squaresize=12pt
\newdimen\thickness \thickness=0.7pt

\def\square#1{\hbox{\vrule width \thickness
   \vbox to \squaresize{\hrule height \thickness\vss
      \hbox to \squaresize{\hss#1\hss}
   \vss\hrule height\thickness}
\unskip\vrule width \thickness} \kern-\thickness}

\def\cut#1{\hbox{\vrule width-1 \thickness
   \vbox to \squaresize{\hrule height-1 \thickness\vss
      \hbox to \squaresize{\hss#1\hss}
   \vss\hrule height-1\thickness}
\unskip\vrule width +4 \thickness} \kern-\thickness}

\def\vsquare#1{\vbox{\square{$#1$}}\kern-\thickness}

\nc{\cre}{\color[rgb]{1.00,0.00,0.00}}
\nc{\cgr}{\color[rgb]{0.00,1.00,0.00}}

\makeatletter
\def\hlinewd#1{%
\noalign{\ifnum0=`}\fi\hrule \@height #1 %
\futurelet\reserved@a\@xhline} \makeatother

\begingroup
\makeatletter\ifx\SetFigFont\undefined
\gdef\SetFigFont#1#2#3#4#5{\reset@font\fontsize{#1}{#2pt}\fontfamily{#3}\fontseries{#4}\fontshape{#5}\selectfont}
\fi
\endgroup

\title{Spin bits at two loops}
\author{S. Bellucci$^a$, P.-Y. Casteill$^a$,  A. Marrani$^{a,b}$,      C.
Sochichiu$^{a,c,d}$\\   \\
$^a${\it  INFN -- Laboratori Nazionali di Frascati,}\\
Via E. Fermi 40, 00044 Frascati, Italy\\
$^b${\it Universit\`{a} degli Studi "Roma Tre"} \\
{\it Dipartimento di Fisica "Edoardo Amaldi",}\\
 Via della Vasca Navale 84, 00146 Roma, Italy\\
$^c${\it Institutul de Fizic\u a Aplicat\u a A\c S,}\\
str. Academiei, nr. 5, Chi\c{s}in\u{a}u, MD2028 Moldova\\
$^d${\it  JINR -- Bogoliubov Lab. Theor. Phys.,}\\
141980 Dubna, Moscow Reg., Russia }

\begin{document}
\maketitle
\begin{abstract}
We consider the Super Yang--Mills/spin system map to construct the
SU(2) spin bit model at the level of two loops in Yang--Mills
perturbation theory. The model describes a spin system with
chaining interaction. In the large $N$ limit the model is shown to
be reduced to the two loop planar integrable spin chain.
\end{abstract}
% ----------------------------------------------------------------
\section{Introduction}
Large $N$ physics \cite{'tHooft:1974jz} gained considerable
interest in recent years (see \cite{Tseytlin:2004xa} for a recent
review and references) due to the AdS/CFT conjecture enlightenment
\cite{Maldacena:1998re,Gubser:1998bc} and, more recently, to the
consideration of various limits for this correspondence
\cite{Berenstein:2002jq,Tseytlin:2003ac,
Beisert:2003ea,Arutyunov:2003rg,Kruczenski:2003gt,Stefanski:2004cw,Bellucci:2004qr}. These
achievements lead to an intensive study of the anomalous dimensions of local gauge
invariant composite operators in $\mathcal{N}=4$ Super Yang--Mills (SYM) model
\cite{Beisert:2003jj}. The major breakthrough in this investigation was the
discovery of integrability of the matrix of anomalous dimensions in the planar
limit, $N\to\infty$ \cite{Minahan:2002ve,Beisert:2004ry}. These results were
extended to two and higher loops \cite{Serban:2004jf,Beisert:2003jb}.

As it is now clear, there is a one-to-one correspondence between one trace
operators in SYM theory and the states in spin chain models. It was enough to
consider the planar limit of SYM theory. If the nonplanar contribution is
considered, the one trace sector is not conserved anymore and one ends up with
trace splitting and joining in the operator mixing \cite{Beisert:2002ff}. Even in
this case one can still consider a one-to-one map between local gauge invariant
operators and a spin system \cite{Bellucci:2004ru,Bellucci:2004qx}. In this case
one has to introduce a set of new degrees of freedom, beyond the spin states,
which describes the chaining state of our spin system. This new field takes values
in the symmetry group of permutations of spin bits and introduces a new gauge
degree of freedom. An alternative approach to the description of the nonplanar
contribution is discussed in \cite{Bellucci:2004fh}.

In this paper we extend the analysis of
\cite{Bellucci:2004ru,Bellucci:2004qx} to the two loop level of
SYM perturbation theory, i.e. we consider the SYM anomalous
dimension/mixing matrix to two loops and apply the map to the spin
bit system to this matrix.

The plan of the paper is as follows. In the next section we
introduce the notations; then in section 3 we consider the two
loop nonplanar anomalous dimension matrix which we map to an
operator acting on the spin bit space. Using the properties of the
symmetry group we are able to reduce the SU(2) nonplanar two loop
Hamiltonian to a remarkably simple form. Finally, in the last
section we draw some conclusions.

In this paper we use conventions and notations of
\cite{Bellucci:2004ru,Bellucci:2004qx}.

% ----------------------------------------------------------------
\section{The setup and the one loop result}
We consider the SU(2) sector of local gauge invariant SYM operators which are
generated by two holomorphic (multi)trace operators built from two complex SYM
scalars $\phi=\phi_5+\ii\phi_6$ and $Z=\phi_1+\ii\phi_2$, with the typical form
\[
  \Op=\tr (\phi Z\phi\phi Z\dots)\tr (\phi\phi\phi Z\dots)\tr(\dots)\dots
\]

This trace can be written in the following explicit form using a permutation group
element $\gamma\in S_L$~:
\bea
  \Op &=&\phi_{i_1}^{a_1a_{\gamma_1}}\phi_{i_2}^{a_2a_{\gamma_2}}\dots\phi_{i_L}^{a_La_{\gamma_L}}\nn\\
  &\equiv& \ket{\phi_{i_1},\ \ldots,\ \phi_{i_L}\ ;\ \gamma},\nn
\eea
where $L$ is the total number of ``letters'' $\phi_i=(\phi,Z)$ in $\Op$ which are
numbered by a label $k=1,\dots, L$. The permutation element $\gamma_k$
gives the next multiplier to the $k$-th letter
$$\gamma\equiv(\gamma_1\ \gamma_2\ldots\gamma_k\ldots\gamma_L)~: \quad \left(
 \begin{array}{cccccc}  1 & 2&\ldots &k &\ldots& L\cr \gamma_1
 & \gamma_2 &\ldots &\gamma_k &\ldots &\gamma_L
 \end{array}\
 \right)\in S_L.$$

 Obviously, the reshuffling of
the labels $k\mapsto \sigma_k$  accompanied by a conjugation of
$\gamma$ with the same group element
$\sigma^{-1}\cdot\gamma\cdot\sigma$ leaves the trace form of $\Op$
unchanged. Therefore, the configurations related by such a
transformation should be considered as equivalent
\begin{equation}
  (\phi_k,\gamma)\sim(\phi_{\sigma_k},\sigma^{-1}\cdot\gamma\cdot\sigma).
\label{equiv} \end{equation} Now, we should map the space of such
operators to the system of $L$ SU(2) $\ft12$-spins (spin bits). The
map is completed by associating to each bit the spin value
$\ket{-\ft12}$,  if we find in the respective place the letter $\phi$,
and $\ket{+\ft12}$, if we find $Z$.

In perturbation theory, the anomalous dimension matrix is given by
\begin{equation}\label{dil0}
 \Delta(g)=\sum_k H_{2k} \lambda^{2k},
 \end{equation}
with $\lambda^2={g_{YM}^2 N \over 8\pi^2 }$  being the 't Hooft
coupling. The coefficients in this expansion are given in terms of
effective vertices, i.e. the operators $H_{2k}$ . They can be
determined e.g. by an explicit evaluation of the divergencies of
two-point function  $\langle \Op(0)\Op(x)\rangle$ Feynman
amplitudes.

At the zero, one and two loop level, the SU(2) anomalous dimension matrices
are given by the following expressions \cite{Beisert:2003tq}:

\begin{align}
 H_0 &=  \tr (\phi \check{\phi}+Z\check{Z}), \nn\\
 H_2 &= -{2\over N} : \tr ([ \phi,Z][\check{\phi},\check{Z}]):,\nn\\
 H_{4}&=
 \frac{1}{N^{2}}\left\{ 2:\tr\left(\left[ Z,\phi \right] \left[ \check{Z},\left[
 Z,\left[ \check{Z},\check{\phi }\right] \right] \right]\right) :\right. +2:\tr\left(\left[
 Z,\phi \right] \left[ \check{\phi },\left[ \phi ,\left[
 \check{Z},\check{\phi }\right] \right] \right]\right) :+  \notag \\
 &\qquad \left. +4N:\tr\left(\left[ \phi,Z \right] \left[
 \check{\phi},\check{Z}\right]\right):\right\}~,\nn
\end{align}
where the checked letters $\check{\phi}$ and $\check{Z}$
correspond to derivatives with respect to the matrix elements
$$\check{Z}_{ij}=\frac{\pd}{\pd Z^{ji}}~,\qquad
  \check{\phi}_{ij}=\frac{\pd}{\pd\phi^{ji}}$$
and colons denote the ordering in which all checked letters in the
group are assumed to stay on the right of the unchecked ones.

In order to find the ``pull back'' of the Hamiltonian \eqref{dil0}
to the spin description, one has to apply it on a (multi)trace
operator corresponding to the spin bit state $\ket{s,\gamma}$ and
map the result back to the corresponding spin bit state. This can
be done term-by-term in the perturbation theory expansion series.

A simple form for the one-loop non-planar Hamiltonian was found earlier
\cite{Bellucci:2004ru,Bellucci:2004qx} (see also \cite{Peeters:2004pt} for a related discussion) and reads
\begin{equation}
H_2=\ft1{2N} \sum_{k,l} H_{k l}\  \Sigma_{ k \gamma_l}=\ft1{N}
\sum_{k,l} (1-P_{kl})\ \Sigma_{ k \gamma_l}, \label{d2}
\end{equation}
where the permutation and chain ``twist" operators are
respectively defined in the following way $(k,l=1,...,L)$:
\begin{eqnarray}
 P_{k l}\,\ket{\{\dots A_k \dots  A_l \dots
 \}}&=&
 \ket{\{\dots A_l \dots  A_k \dots
 \}}, ~{\rm with }~A_l,A_k\in\{\phi,Z\}\nn\\
 \label{sigma}
 \Sigma_{k l}\ket{\gamma}&=&
 \begin{cases}
 \ket{\gamma\,\sigma_{kl}}&\text{if }k \neq l
 \\
 N\ket{\gamma},& k=l.
 \end{cases}
\end{eqnarray}

\begin{figure}[t]
            \begin{center}
               \scalebox{0.8}{
               {\begin{picture}(0,0)%
\epsfig{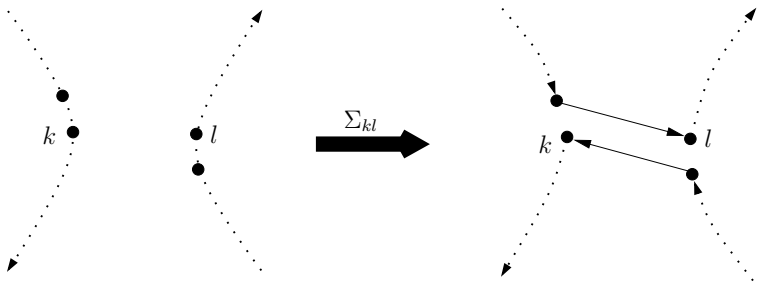}%
\end{picture}%
\setlength{\unitlength}{3947sp}%
\begin{picture}(5929,2204)(3776,-6132)
\put(6451,-4884){\makebox(0,0)[lb]{\smash{\SetFigFont{12}{14.4}{\rmdefault}{\mddefault}{\updefault}
$\Sigma_{kl}$
}}}
\put(5401,-5002){\makebox(0,0)[lb]{\smash{\SetFigFont{12}{14.4}{\rmdefault}{\mddefault}{\updefault}
$l$
}}}
\put(9280,-5040){\makebox(0,0)[lb]{\smash{\SetFigFont{12}{14.4}{\rmdefault}{\mddefault}{\updefault}
$l$
}}}
\put(4194,-5012){\makebox(0,0)[rb]{\smash{\SetFigFont{12}{14.4}{\rmdefault}{\mddefault}{\updefault}
$k$
}}}
\put(8088,-5081){\makebox(0,0)[rb]{\smash{\SetFigFont{12}{14.4}{\rmdefault}{\mddefault}{\updefault}
$k$
}}}
\end{picture}}}
               \caption{Splitting and joining of chains by $\Sigma_{kl}$.}\label{1in2}
            \end{center}
\end{figure}

$\Sigma_{kl}$ acts as a chain splitting and joining operator as
illustrated in Figure \ref{1in2}. Notice that two $\Sigma$'s do not commute
if they have indices in common. The factor $N$ in the case $k=l$
in Eq. \eqref{sigma} appears because the splitting of a trace at
the same place leads to a chain of length zero, whose
corresponding trace is ${\tr}\, 1=N$. It is important to note that
the operator $\Sigma_{kl}$ acts only on the linking variable,
while the the two-site SU(2) one-loop spin bit Hamiltonian $H_{k
l}=2\ (1-P_{kl})$ acts on the spin space. Therefore, the two
operators commute.

% ----------------------------------------------------------------
\section{The two loop Hamiltonian}

Let us now consider the two loop Hamiltonian \bea H_{4}&=
 \frac{1}{N^{2}}\left\{ 2:\tr(\left[ Z,\phi \right] \left[ \check{Z},\left[
 Z,\left[ \check{Z},\check{\phi }\right] \right] \right]) :\right. +2:\tr(\left[
 Z,\phi \right] \left[ \check{\phi },\left[ \phi ,\left[
 \check{Z},\check{\phi }\right] \right] \right]) :+  \notag \\
 &\qquad \left. +4N:\tr(\left[ \phi,Z \right] \left[
 \check{\phi},\check{Z}\right]):\right\}.
\label{d4}
\eea

We introduce the operator
$${\cal O}_{B_1,B_2,B_3}=\tr(
\check{A}_k\  A_{B_1}\ \check{A}_l\ A_{B_1}\
\check{A}_m \ A_{B_3} )~,$$
with
$\check{A}_k,\check{A}_l,\check{A}_m=\check{\phi},\check{Z}$
acting on the $k,l,m$-th sites of the state
$|...A_k...A_l...A_m...\ ;\gamma\rangle$, respectively. Here $B_i$
are non-intersecting sequences chosen in the set $\{klm\}$ and $A_{B_i}$ are then
monomials in $A_k,A_l,A_m$. For example, the choices $B_1=\emptyset,k,lm,klm$
correspond to $A_{B_1}=1,A_k,A_lA_m,A_kA_lA_m$. Indeed, as $A_{B_1},A_{B_2},A_{B_3}$
are made of the same number of
$\phi$ and $Z$ than in $\check{A}_k,\check{A}_l,\check{A}_m$, any
trace of (\ref{d4}) can be written with such an ${\cal
O}_{B_1,B_2,B_3}$ operator.

Acting with ${\cal O}_{B_1,B_2,B_3}$ on a spin chain state
specified by $\gamma$, one finds \bea
{\cal O}_{B_1,B_2,B_3}:\gamma=\left(
\begin{array}{llllll}
  \gamma^{-1}_k & k & \gamma^{-1}_l & l &  \gamma^{-1}_m & m\\
  k & \gamma_k &  l & \gamma_l &  m & \gamma_m  \\
\end{array}
\right)\to
\left(
\begin{array}{llllll}
  \gamma^{-1}_k & B_1 & \gamma^{-1}_l & B_2 &  \gamma^{-1}_m & B_3\\
  B_1 & \gamma_l &   B_2 &\gamma_m &   B_3 & \gamma_k  \\
\end{array}
\right).\nn \eea
\begin{figure}[h]
            \begin{center}
               \scalebox{0.7}{
               \begin{picture}(0,0)%
\epsfig{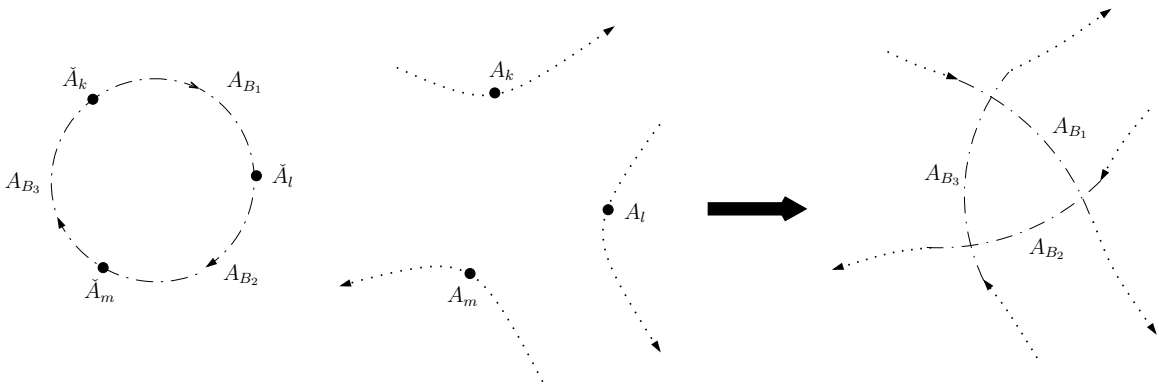}%
\end{picture}%
\setlength{\unitlength}{3947sp}%
\begin{picture}(10350,3409)(76,-6261)
\put(8326,-4411){\makebox(0,0)[lb]{\smash{\SetFigFont{12}{14.4}{\rmdefault}{\mddefault}{\updefault}
$A_{B_3}$
}}}
\put(4418,-3488){\makebox(0,0)[lb]{\smash{\SetFigFont{12}{14.4}{\rmdefault}{\mddefault}{\updefault}
$A_k$
}}}
\put(5634,-4766){\makebox(0,0)[lb]{\smash{\SetFigFont{12}{14.4}{\rmdefault}{\mddefault}{\updefault}
$A_l$
}}}
\put(4045,-5514){\makebox(0,0)[lb]{\smash{\SetFigFont{12}{14.4}{\rmdefault}{\mddefault}{\updefault}
$A_m$
}}}
\put(590,-3571){\makebox(0,0)[lb]{\smash{\SetFigFont{12}{14.4}{\rmdefault}{\mddefault}{\updefault}
$\check{A}_k$
}}}
\put(2472,-4445){\makebox(0,0)[lb]{\smash{\SetFigFont{12}{14.4}{\rmdefault}{\mddefault}{\updefault}
$\check{A}_l$
}}}
\put(2065,-3581){\makebox(0,0)[lb]{\smash{\SetFigFont{12}{14.4}{\rmdefault}{\mddefault}{\updefault}
$A_{B_1}$
}}}
\put(2034,-5312){\makebox(0,0)[lb]{\smash{\SetFigFont{12}{14.4}{\rmdefault}{\mddefault}{\updefault}
$A_{B_2}$
}}}
\put(788,-5478){\makebox(0,0)[lb]{\smash{\SetFigFont{12}{14.4}{\rmdefault}{\mddefault}{\updefault}
$\check{A}_m$
}}}
\put( 76,-4486){\makebox(0,0)[lb]{\smash{\SetFigFont{12}{14.4}{\rmdefault}{\mddefault}{\updefault}
$A_{B_3}$
}}}
\put(9474,-3991){\makebox(0,0)[lb]{\smash{\SetFigFont{12}{14.4}{\rmdefault}{\mddefault}{\updefault}
$A_{B_1}$
}}}
\put(9264,-5079){\makebox(0,0)[lb]{\smash{\SetFigFont{12}{14.4}{\rmdefault}{\mddefault}{\updefault}
$A_{B_2}$
}}}
\end{picture}
}
               \caption{The action of ${\cal O}_{B_1,B_2,B_3}$ on a spin chain state.}\label{fig:mafigureklm}
            \end{center}
\end{figure}
A pictural view of the action of ${\cal O}_{B_1,B_2,B_3}$ is given
in Figure \ref{fig:mafigureklm}.
The three relevant cases are
 \bea
{\cal O}_{klm,\emptyset,\emptyset}\,\ket{\gamma}&=&\ket{\footnotesize\left(
\begin{array}{llllll}
  \gamma^{-1}_k & k &l&m& \gamma^{-1}_l  &  \gamma^{-1}_m \\
  k & l& m& \gamma_l &   \gamma_m  & \gamma_k
  \\
\end{array}
\right)}=P_{lm}\,\Sigma_{l \gamma_m }\,\Sigma_{\gamma_k m}\,\ket{\gamma}\nn\\
%\Sigma_{\gamma_l \gamma_m}\,\Sigma_{l \gamma_m }\,\Sigma_{\gamma_k m}\,\Sigma_{lm} \gamma\nn\\
{\cal O}_{m,\emptyset,kl}\,\ket{\gamma}&=&\ket{\left(\footnotesize
\begin{array}{llllll}
  \gamma^{-1}_k & m & \gamma^{-1}_l  &  \gamma^{-1}_m & k&l\\
  m & \gamma_l  &\gamma_m &   k&l & \gamma_k  \\
\end{array}%
\right)}=P_{km} \Sigma_{l\gamma_m} \Sigma_{\gamma_k\gamma_l}\,\ket{\gamma} \nn\\
{\cal O}_{kl,\emptyset,m}\,\ket{\gamma}&=&\ket{\left(\footnotesize
\begin{array}{llllll}
  \gamma^{-1}_k & k&l & \gamma^{-1}_l  &  \gamma^{-1}_m & m\\
  k&l & \gamma_l  &\gamma_m &   m & \gamma_k  \\
\end{array}
\right)}=P_{km}\Sigma_{km}\Sigma_{l\gamma_m}\,\ket{\gamma}.\label{os}
\eea
In order to write the operators in terms of $P$ and $\Sigma$, we used the fact that permutations
can also be viewed as operators acting on $\gamma$ rather than on
spin states. From such a viewpoint, the action of $P_{kl}$ on
$\gamma$ is
$$P_{kl} \ket{\footnotesize\left(
\begin{array}{lllllllll}
. & . & . & k & . & l & . & . & .\\
. & k & . & . & . & . & . & l & .
\end{array}
\right)}~ = ~\ket{\footnotesize\left(
\begin{array}{lllllllll}
. & . & . & l & . & k & . & . & .\\
. & l & . & . & . & . & . & k & .
\end{array}
\right)}~,$$
while the action of $\Sigma_{kl}$ is given explicitly by
$$\Sigma_{kl} \ket{\footnotesize\left(
\begin{array}{lllll}
. & . & . & . & .\\
. & k & . & l & .
\end{array}
\right)} ~=~ \ket{\footnotesize\left(
\begin{array}{lllll}
. & . & . & . & .\\
. & l & . & k & .
\end{array}
\right)}~.$$

All other contributions in (\ref{d4}) can be written as permutations
and/or relabelling of indices in \eqref{os}. Collecting the sixteen
terms coming from \eqref{d4}, one finds
\bea H_4 &=&{2\over N^2}
\, \sum_{k\neq l\neq m}\left[ (P_{km}
P_{lm}+P_{kl}P_{lm}-P_{kl}-P_{km})(\Sigma_{l\gamma_m}\,\Sigma_{\gamma_k\gamma_l}
+
\Sigma_{kl}\,\Sigma_{\gamma_l m})\right.\nn\\
&&\left.+(2P_{lm}+2P_{kl}-2-P_{kl} P_{lm}-P_{lm}P_{kl}
) \Sigma_{k\gamma_l}\,\Sigma_{l\gamma_m} \right]\nn\\
&& +{4\over N} \sum_{k,l} (P_{kl}-1)\Sigma_{k \gamma_l}.
\label{d40} \eea Using the relation%s
\bea
1-P_{kl}-P_{km}-P_{lm}+P_{km} P_{lm}+P_{kl}P_{lm}=0 %\\
%1-P_{kl}-P_{km}-P_{lm}+P_{kl} P_{lm}+P_{lm}P_{kl}=0
\label{identity-1loop}\eea
valid for $k\neq l\neq m$ on $\ft12$-spin  states,
 one can rewrite the two loop SU(2) spin bit Hamiltonian (\ref{d40}) as \bea H_4
&=&{2\over N^2} \, \sum_{k\neq l\neq m}\left[
(P_{lm}-1)(\Sigma_{l\gamma_m}\,\Sigma_{\gamma_k\gamma_l}+
\Sigma_{kl}\,\Sigma_{\gamma_l m})+(P_{lm}+P_{kl}-P_{km}-1) \Sigma_{k\gamma_l}\,\Sigma_{l\gamma_m} \right]\nn\\
&& +{4\over N} \sum_{k,l} (P_{kl}-1)\Sigma_{k \gamma_l}.
\label{d4bis} \eea

\begin{figure}[h]\label{fig:klm}
            \begin{center}
               \scalebox{0.8}{
               \begin{picture}(0,0)%
\epsfig{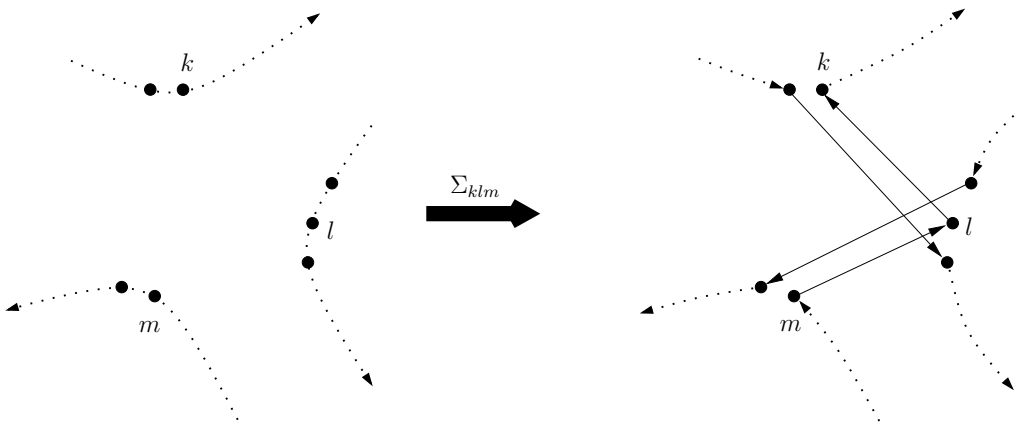}%
\end{picture}%
\setlength{\unitlength}{3947sp}%
\begin{picture}(7973,3298)(2895,-6375)
\put(9302,-3577){\makebox(0,0)[lb]{\smash{\SetFigFont{12}{14.4}{\rmdefault}{\mddefault}{\updefault}
$k$
}}}
\put(10465,-4878){\makebox(0,0)[lb]{\smash{\SetFigFont{12}{14.4}{\rmdefault}{\mddefault}{\updefault}
$l$
}}}
\put(6421,-4546){\makebox(0,0)[lb]{\smash{\SetFigFont{12}{14.4}{\rmdefault}{\mddefault}{\updefault}
$\Sigma_{klm}$
}}}
\put(5453,-4908){\makebox(0,0)[lb]{\smash{\SetFigFont{12}{14.4}{\rmdefault}{\mddefault}{\updefault}
$l$
}}}
\put(4305,-3585){\makebox(0,0)[lb]{\smash{\SetFigFont{12}{14.4}{\rmdefault}{\mddefault}{\updefault}
$k$
}}}
\put(3977,-5642){\makebox(0,0)[lb]{\smash{\SetFigFont{12}{14.4}{\rmdefault}{\mddefault}{\updefault}
$m$
}}}
\put(9011,-5627){\makebox(0,0)[lb]{\smash{\SetFigFont{12}{14.4}{\rmdefault}{\mddefault}{\updefault}
$m$
}}}
\end{picture}
}
               \caption{Splitting and joining of chains by $\Sigma_{klm}\equiv\Sigma_{k\gamma_l}\Sigma_{l\gamma_m}$.}\label{1in3}
            \end{center}
\end{figure}

  Next, we would like to write the Hamiltonian in terms of $P$'s and a single (two loop) joining-splitting operator
$$\Sigma_{klm}\equiv \Sigma_{k\gamma_l}\,\Sigma_{l \gamma_m}.$$
Its action is
depicted in Figure \ref{1in3}.
Because a one-operator trace vanishes
and a two-operators trace is fully symmetric, the joining-splitting $\Sigma_{klm}$ operator satisfies
\begin{equation}
\Sigma_{llm}=\Sigma_{kll}=(P_{kl}-1)\ \Sigma_{klk}=0~.
\label{van}
\end{equation}
Noticing that\footnote{Here $k\neq l\neq m$ is
understood.}
$$\begin{array}{llll}
\Sigma_{l\gamma_m}\,\Sigma_{\gamma_k\gamma_l} &=\Sigma_{\gamma_k\gamma_l}\,
\Sigma_{l \gamma_m} &= \Sigma_{\gamma_k l m} & {\rm for}~l\neq \gamma_k\\
\Sigma_{kl}\,\Sigma_{\gamma_l m} &=\Sigma_{m\gamma_l}\, \Sigma_{l k} &= \Sigma_{m l
\gamma^{-1}_k}\qquad ~~~& {\rm for}~k\neq \gamma_l~,
\end{array}$$
the first term of the first sum in (\ref{d4bis}) can then be rewritten as
 \bea
  X&\equiv& \sum_{k\neq l\neq m}(P_{lm}-1)(\Sigma_{l\gamma_m}\,\Sigma_{\gamma_k\gamma_l}+
\Sigma_{kl}\,\Sigma_{\gamma_l m})\nn\\ \nn\\
&=&\sum_{\gamma^{-1}_{k'}\neq l\neq m} \Sigma_{k' l m}(P_{lm}-1)+
 \sum_{k\neq l\neq \gamma_{m'}} \Sigma_{k l m'}(P_{kl}-1)\nn\\\nn\\
 &&+  \, \sum_{l,m}
(P_{lm}-1)(\Sigma_{l\gamma_m}\,\Sigma_{l\gamma_l}+\Sigma_{\gamma_l
l}\,\Sigma_{\gamma_l m})~,\nn
\eea
where the terms with $l=\gamma_k$ or $k=\gamma_l$ were explicitly
written apart. Relaxing the restrictions in the sums by
subtracting the exceptional terms and using $\Sigma_{kk}=N {\cal
I}$, one gets
\bea
X&=&\sum_{k,l,m} \Sigma_{k l m}(P_{lm}+P_{kl}-2)-2
N\,\sum_{l,m}(P_{lm}-1) \Sigma_{l \gamma_m} \nn\\\nn\\
 &&-
\sum_{l,m}(P_{lm}-1)\left[ \Sigma_{\gamma_m \gamma_l}
\Sigma_{l\gamma_m}-
 \Sigma_{l\gamma_m}\Sigma_{l\gamma_l}
 +\Sigma_{m \gamma_l}\Sigma_{l m}-\Sigma_{\gamma_l l}\,\Sigma_{\gamma_l
 m}\right]~.\nn
\eea
Finally,  using the relations $\Sigma_{\gamma_m\gamma_l}\Sigma_{l
 \gamma_m}=\Sigma_{l \gamma_m}\Sigma_{l \gamma_l}$ and $\Sigma_{m \gamma_l}\Sigma_{l
m}=\Sigma_{\gamma_l l}\Sigma_{\gamma_l m}$
and plugging $X$ in (\ref{d4bis}), one finally finds the
surprisingly simple result of this paper
\begin{equation}
 H_4\,=\,{2\over N^2} \, \sum_{k,l,m} (2 P_{lm}+2 P_{kl}-P_{km}-3)
 \Sigma_{klm}. \label{finald4}
\end{equation}
Notice that in (\ref{finald4}) one can chose to put or not the restrictions $k\neq l\neq m$, as equalling two indices
always gives a term of the form (\ref{van}).

It is also instructive to rewrite the Hamiltonian in terms of SU(2) spin operators
$\vec{s}=\ft12\vec{\sigma}$, where $\vec{\sigma}$ are usual Pauli matrices, using the identity (see e.g. \cite{Faddeev:1996iy})

\begin{equation}
  P_{kl}=\ft12+\ft12\vec{s}_k\cdot\vec{s}_l.
\end{equation}
After the substitution of permutation operators, the one-loop Hamiltonian \eqref{d2} takes the following
form~:
\begin{multline}
  H_4=
  {8\over N^2}
 \sum_{k,l,m}((\vec{s}_k-2\vec{s}_l+\vec{s}_m))^2
 \Sigma_{klm}
 ={8\over N^2}
 \sum_{k,l,m}((\vec{s}_k-\vec{s}_l)-(\vec{s}_l-\vec{s}_m))^2
 \Sigma_{klm}.
\end{multline}
This Hamiltonian has a simple meaning (see figure \ref{fig:klm}): $\Sigma_{klm}$
cyclically exchanges the incoming and outgoing ends of the chains adjacent to the bits
$k,l$ and $m$; at the same time the spin part acts as the discrete second
derivative along the new chain.  After knowing that the one-loop Hamiltonian has
the similar structure
\[
  H_{2}=\frac{4}{N}\sum_{k,l}(\vec{s}_{k}-\vec{s}_l)^2
  \,\Sigma_{k\gamma_l},
\]
it is very tempting to conjecture that at the arbitrary $n$-loop level the Hamiltonian
is given by the discrete derivative of the order $n$ squared times the splitting
that cyclically exchange the chain ends~:
\begin{equation*}
  H_{n}\sim\frac{4n}{N^n}\sum_{k_1,\dots,k_n}
  \left(\sum_{i=0}^n \frac{n!}{(n-i)!i!}(-1)^i \vec{s}_{k_i}\right)^2
  \,\Sigma_{k_0\gamma_{k_1}}\Sigma_{k_1\gamma_{k_2}}\dots\Sigma_{k_{n-1}\gamma_{n}}.
\end{equation*}
This is compatible with the BMN conjecture \cite{Berenstein:2002jq} but it implies
that the Hamiltonian can be written linearly in pair permutation operators $P_{kl}$
at \emph{any loop level}, which unfortunately is probably not the case.

\subsubsection*{The planar limit}  $N\rightarrow\infty$ affects just the ``twist"
operator in the following way\footnote{In fact from Eq. \eqref{sigma} the
following decomposition of $\Sigma_{kl}$ holds:
$$
 \Sigma_{kl}=N\delta_{kl}+(1-\delta_{kl}){\widetilde{\Sigma}\\}_{kl},
$$
where ${\widetilde{\Sigma}\\}_{kl}$ is the joining-splitting operator spoiled of its degeneracy in the case of
coinciding sites.}:
$$\lim_{N\rightarrow \infty }\ {1\over N}
\Sigma_{kl}=\delta_{kl}.$$
The two loop non-planar SU(2) spin bit
Hamiltonian \eqref{finald4} gives then the correct
known expression \cite{Beisert:2003tq} in the planar limit
$N\rightarrow\infty$\footnote{We assume here a single trace, so that $\gamma_{k}\equiv k+1$,
with the identification $L+1\equiv 1$.}
\bea
 \lim_{N\rightarrow \infty   }H_{4} &=&
 2\sum_{k=1}^{L}\left(4P_{k,k+1}-P_{k,k+2} -3\right)\nn\\
 &=&
 2\sum_{k=1}^{L}\left( -4+6P_{k,k+1}-P_{k,k+1}\
 P_{k+1,k+2}-P_{k+1,k+2}\
 P_{k,k+1}\right),\label{planard4}
\eea
where in the last passage we used the identity \eqref{identity-1loop}.

\section{Conclusion}
In this paper we considered the anomalous dimension and mixing of composite
operators of $\mathcal{N}=4$ SYM theory in the SU(2) symmetric
sector. Using the isomorphic map between gauge invariant composite operators and
the spin bit states, we computed the spin bit Hamiltonian corresponding to two-loop
corrections to the anomalous dimension/mixing matrix. The resulting Hamiltonian at
this level has two important properties:
\begin{itemize}
  \item[\emph{i})] The Hamiltonian shows  at the two-loop level an explicit full
factorization in the spin and chain splitting parts similar to the
one-loop level.

  \item[\emph{ii})] Its action is given by a three-point spin interaction and a cyclic
exchange (hopping) of the chain ends.
\end{itemize}

The first property is expected to hold at any loop order since the Hilbert
space of the spin bit model is always the direct product of the spin space
and the linking variable $\gamma$-space.
The second property has a natural generalization to $n+1$
interacting points appearing at $n$ loops: cyclic exchange of the chain ends
multiplied by the square of the $n$-th discrete derivative of spin
operators $\vec{s}_k$. In the continuum limit, this is in perfect agreement with the BMN conjecture which gives
a term $\sim \lambda^{2n} (\pd^n\phi)^2$ as the $n$ loop contribution .

A strong consequence of
this higher loop conjecture is the requirement that at
any loop level the spin part of the Hamiltonian should always be linear in
permutation operators. This implies strong restrictions on the planar limit too. Of course,
there is a very rich set of identities involving permutation operators which
could be used to prove such a property. Our attempts to
check this at the three-loop level with the expressions for planar
Hamiltonians given by \cite{Beisert:2003tq} so far failed.

Similar results giving the Hamiltonian at three and more loops in terms of spin-bit would give more insight, allowing one to
give a conjecture for generalization. In fact, there is enough data
and technique at this stage to produce the three loop Hamiltonian. The problem being only
algebraic difficulty, it seems hopefully superable by the use of computer algebra.

Finally, we notice that it would be interesting to extend this analysis to other sectors of
$\mathcal{N}=4$ SYM ; unfortunately, only SU(2) anomalous dimension
operators are known beyond one-loop .

\section*{Acknowledgements}

We would like to thank Francisco Morales for useful discussions and collaboration on the early stage of this work.

This work was partially supported by NATO Collaborative Linkage Grant PST.CLG.
97938, INTAS-00-00254 grant, RF Presidential grants MD-252.2003.02,
NS-1252.2003.2, INTAS grant 03-51-6346, RFBR-DFG grant 436 RYS 113/669/0-2, RFBR
grant 03-02-16193 and the European Community's Human Potential Programme under
contract HPRN-CT-2000-00131 Quantum Spacetime.
% ----------------------------------------------------------------
\bibliographystyle{utphys}
\providecommand{\href}[2]{#2}\begingroup\raggedright\endgroup

% ---------------------------------------------------------------
\end{document}